# Uniaxial-stress induced phase transitions in $[001]_C$-poled $0.955Pb(Zn_{1/3}Nb_{2/3})O_3$-$0.045PbTiO_3$

Pierre-Eymeric Janolin and Brahim Dkhil

*Laboratoire Structures, Propriétés et Modélisation des Solides, CNRS UMR 8580, Ecole Centrale Paris, 92295 Châtenay-Malabry Cedex, France*

Matthew Davis, Dragan Damjanovic, and Nava Setter

*Ceramics Laboratory, Swiss Federal Institute of Technology-EPFL, 1015 Lausanne, Switzerland*

ABSTRACT

First-order, rhombohedral to orthorhombic, stress-induced phase transitions have been evidenced by bulk charge-stress measurements and X-ray diffraction derived lattice strain measurements in $[001]_C$-poled PZN-4.5PT. The transitions are induced by uniaxial, compressive loads applied either along or perpendicular to the poling direction. In each case, they occur via rotation of the polar vector in the *Cm* monoclinic plane and the induced lattice strain is hysteretic yet reversible. Although no depoling is observed in the transverse mode, net depolarization is observed under longitudinal stress which is important for the use of [001]c-poled PZN-PT and PMN-PT single crystals in Tonpilz-type underwater projectors.

Relaxor-based ferroelectric single crystals, (1-x)Pb(Zn$_{1/3}$Nb$_{2/3}$)O$_3$-xPbTiO$_3$ [PZN-xPT] and (1-x)Pb(Mg$_{1/3}$Nb$_{2/3}$)O$_3$-xPbTiO$_3$ [PMN-xPT] exhibit extraordinarily high electromechanical properties that make them good candidates for next-generation sensor, medical imaging and transducer applications[1-3]. Electric-field induced phase transitions have been extensively studied in PZN-xPT and PMN-xPT by high resolution diffraction[4-6] and polarized light microscopy[7,8], and, macroscopically, by strain-field measurements[1,9]. However, much less work has been carried out on their response to mechanical stresses[10,11], and the phase transitions they can induce[9]. Notably, the direct piezoelectric (charge-stress) response of PMN-xPT and PZN-xPT is directly relevant to their use in sensor applications[11]; in Tonpilz-type underwater projectors, uniaxial stress is required to ensure that the piezoelectric elements remain in compression at all times[10].

The purpose of this letter is to investigate the effect of uniaxial stress on the structure, and electric and mechanical properties of domain-engineered, single crystal PZN-4.5PT [$x = 4.5\%$]. When rhombohedral PZN-4.5PT is cut and poled along a non-polar [001]$_C$ (pseudocubic) direction the resultant "4R", domain-engineered state is composed of four degenerate, domain variants each with <111>$_C$-type polar vectors equally inclined to the poling direction [see fig. 1(a)][3]. Stress-induced phase transitions are evidenced both macroscopically, by charge-stress measurements, and by X-ray diffraction (XRD) when compressive stresses are applied both along and perpendicular to the poling direction (longitudinal and transverse modes). In the longitudinal mode, depoling is observed above a critical stress as a direct result of the stress-induced phase transition; in contrast, no depoling is observed under transverse loads. The reason for this behavior, and its importance to the use of domain-engineered PZN-xPT and PMN-xPT in sensor applications, is discussed.

Charge-stress measurements were made in a Berlincourt-type press, as described elsewhere[9,11]. Quasi-static charge-stress loops were measured by loading and unloading to compressive stresses between 0 and 200 MPa. Each load/unload cycle took around 2 minutes. In the longitudinal mode [fig. 1(a)], stresses $\sigma_3$ are applied along the poling direction (across its electroded faces) between two flat, well-aligned, steel plates; the change in charge density across the crystal $D_3$ is recorded. In the transverse mode [fig. 1(c)], the sample is turned on its side, a stress $\sigma_1$ is applied perpendicular to the poling direction, and $D_3$ is again measured. $D_3(\sigma_3)$ measurements give a measure of the longitudinal piezoelectric coefficient $d_{33}$; transverse $D_3(\sigma_1)$ measurements give a measure of $d_{31}$.

X-ray diffraction (XRD) measurements were performed, on the same samples, using a high-resolution diffractometer equipped with a 18kW rotating anode generator. Static, uniaxial, compressive stresses (both longitudinal and transverse) were applied *in situ*, again between two flat well-aligned plates, in an in-house built, mechanical press mounted on the diffractometer.

At each load, the stress-induced lattice strain, in the direction perpendicular to the direction of applied stress, was determined by monitoring the shift in the peak position of the (040), (240) and (340) reflections. In the transverse (longitudinal) mode, the stress $\sigma_1$ ($\sigma_3$) is applied perpendicular to (along) the $[001]_C$ poling direction and the lattice strain $S_3$ ($S_1$) is measured along (perpendicular to) the poling direction.

For the transverse measurements, <001>$_C$-oriented single crystals of PZN-4.5PT were purchased from TRS Ceramics (State College, PA) with nominal dimensions of $4 \times 4 \times 2$ mm$^3$. Gold electrodes were sputtered onto the two largest surfaces of the crystals. For the longitudinal measurements, the crystals were cut into two 2 x 2 x 4 mm$^3$ columnar samples and gold electrodes were sputtered onto the two smallest 2 x 2 mm$^2$ faces. Aspect ratios for the measurements were chosen, based on previous work, to minimize the effects of clamping[9,11]. The crystals were poled by applying a small field (200 V/mm) whilst cooling from the cubic phase to room temperature.

Measured charge-stress loops in both transverse and longitudinal modes [figures 2(a) and (b), respectively] were hysteretic and evidence of a first order, stress-induced phase transition, as already noted in longitudinally-loaded, $[001]_C$-poled PMN-xPT and PZN-xPT[9,12,13].

In the transverse mode [figure 2(a)], there is an initial, linear and anhysteretic segment below a threshold stress $\sigma_T \approx 20$ MPa; the gradient of the linear portion corresponds to a transverse piezoelectric coefficient $d_{31} = -1290$ pC/N. Above the threshold stress, a kink in the curve and a hysteresis is evidence of a first-order phase transition. At higher stresses (> 30 MPa), there is a second, anhysteretic, nearly-linear segment with a gradient ($d_{31}$) of around −170 pC/N. Moreover, the loop is completely closed, indicating its complete reversibility, with zero depolarization ($\Delta D_3 = 0$). Most notably, the form of the curve is very similar to that seen in converse (strain-field) measurements of electric-field induced phase transitions in [001]$_C$-oriented PMN-xPT and PZN-xPT[3,9].

Note that in [001]$_C$-poled, 4R PZN-4.5PT there is no driving force for ferroelastic domain switching when a compressive stress is applied either along, or perpendicular to, the poling direction[11] [figures 1(a) and (c)]. Charge-stress loops indicative of ferroelastic switching, for example in non-domain engineered crystals[9] and ceramics[14], are characteristically more hysteretic and not closed since the process is irreversible leading to net depolarization.

Figure 3(a) shows the stress evolution of lattice strain $S_3$ under transverse loading. The strain-stress loop is hysteretic and completely closed, without remanent strain when the sample is unloaded. The first part of the cycle, below a stress of around 20 MPa, is linear; it is then hysteretic between 20 and 30 MPa, before becoming linear again at higher stresses. The Young's modulus $Y$ calculated from the gradient of the initial linear segment is 12 GPa; it is around 30 GPa for the second, high-stress part.

Unfortunately, the crystal structure could not be determined due to the restrictions imposed on the X-ray beam by the sample holder. However, figure 3(a) is strong evidence that the charge-stress loop shown in figure 2(a) is indeed intrinsic in origin; that is, it is consistent with a stress-induced phase transition from the rhombohedral phase.

Due to the geometry of the loading conditions[9], and as predicted by first principles calculations[15], the polarization is expected to rotate within a $\{110\}_C$ type (monoclinic[2]) plane towards the $<110>_C$ direction of an orthorhombic phase. The hysteretic jump in strain [fig. 3(a)] and polarization [fig. 2(a)] therefore correspond to a rhombohedral - orthorhombic, first order phase transition via an intermediate $Cm$ monoclinic phase[2]. In this way, the gradient ($-170$ pC/N) of the high-stress portion of figure 2(a) correspond to $d_{32}(\neq d_{31})$ of the two-variant orthorhombic phase [fig. 1(d)].

In the longitudinal mode, strain-stress cycles display qualitatively the same behavior [see figure 3(b)], again indicative of a first-order, stress-induced phase transition. Both loops are closed; there is no remanent lattice strain upon removal of the stress. This time, however, the initial kink occurs at a lower compressive stress of around 10 MPa. Moreover, although the Young's modulus of the high-stress region is close to that in the transverse mode (30 GPa), it is much smaller for the initial linear portion (4 GPa).

As shown in figure 1, the geometry of the loading conditions for each individual rhombohedral domain is identical in both transverse and longitudinal modes. Thus we might expect the same stress-induced phase transition to an orthorhombic phase, via polarization rotation in a $Cm$ $\{110\}_C$ plane. This time, however, the stress-induced domain structure consists of four variants [fig. 1(b)].

Hysteretic behavior is also observed in the longitudinal charge-response of $[001]_C$-poled PZN-4.5PT [fig. 2(b)]. Below a certain stress, the form of the loops was similar to those observed for transverse loading [fig. 2(a)], characteristic of a stress-induced phase transition, and with no depolarization ($\Delta D_3 = 0$). The gradient of the initial, quasi-linear portion of the curve corresponds to a longitudinal piezoelectric coefficient, for polarization rotation in the *Cm* monoclinic plane, of $d_{33} = 1550$ pC/N.

Confusingly, however, the initial kink in the charge-stress loop occurs at a threshold stress $\sigma_T \approx 20$ MPa, which does not coincide with the kink in lattice strain in figure 3(b) ($\sigma_T \approx 10$ MPa). The reason for this is unclear, although it may be related to the fact that whereas charge-stress measurements average the transformation for the entire sample, XRD measurements only sample a small volume (a few microns thick) at the edge of the sample[16]; in this region, the stress state may be different to that in the bulk. Of course, this would not explain why the observed threshold stresses are much closer ($\sigma_T \approx 20$ MPa) in the transverse mode [figures 2(a) and 3(a)]. However, the differing aspect ratios between the two cases might play a role. In the transverse mode, the aspect ratio is chosen such that the piezoelectric contribution from unwanted lateral stresses, due to friction at the sample surfaces, roughly cancel out[11,17]; this is not the case in the longitudinal mode.

At larger stresses (above around 10 MPa), longitudinal charge-stress loops were no longer closed and net depolarization was observed upon unloading [fig. 2(b)]; for repeated cycles, the depolarization $\Delta D_3$ was always close to $-0.5 \times 10^4$ pC/mm². Notably, at very large compressive stresses ($\sigma > 100$ MPa), the inverse gradient tended to zero. This is consistent with the presence of an orthorhombic phase for which $d_{22} = 0$.

The fact that depolarization is observed under large, longitudinal loads, but never in the transverse mode, is an important result; it will be relevant to the use of such crystals in Tonpilz-type projectors which are necessarily subject to uniaxial stress. We postulate that, in the transverse mode, the crystal remains domain-engineered throughout the phase transition. As shown in [fig. 1(d)], the two resultant, orthorhombic variants are equally inclined to the [001]$_C$ poling direction. In contrast, in the longitudinal mode, the domain-engineered structure is broken by the phase transition [see figure 1(b)]. Thus, when the sample is unloaded, rhombohedral domains can also nucleate and grow with polar vectors directed away from the [001]$_C$-poling direction (i.e. "down" domains). This would lead to net depolarization of the sample, but no remanent lattice strain. As noted elsewhere[17], this might also explain the hysteresis reported in the longitudinal, direct piezoelectric response of [001]$_C$-poled PZN-xPT and PMN-xPT under small, dynamic loads (< 10 MPa), and its absence in the equivalent transverse mode[11].

Finally, the possibility also remains that the differing behaviors are due to anisotropy in the polarization rotation mechanism[11]. In the transverse mode, rotation is *toward* the poling direction; in the longitudinal mode it is *away*. This might also explain why the low-stress Young's modulus measured in the longitudinal mode (4 GPa) is lower than that in the transverse mode (12 GPa).

In summary, stress-induced phase transitions have been evidenced by bulk charge-stress measurements and XRD-derived lattice strain measurements in [001]$_C$-poled, rhombohedral PZN-4.5PT, under both transverse and longitudinal, compressive loads. An orthorhombic phase is likely induced via polarization rotation in the *Cm* monoclinic plane and a first-order (hysteretic) phase transition. The phase transition is reversible in both cases. However, although no depoling is observed under transverse stress, net depolarization is observed in the longitudinal mode. This is due to the fact that, in the longitudinal case, the domain-engineered structure is broken by the stress-induced phase transition; this is not the case in the transverse mode. This is an important result for the use of [001]$_C$-poled PZN-xPT and PMN-xPT single crystals in Tonpilz-type projector applications.

M.D., D.D. and N.S. acknowledge funding from the Swiss National Science Foundation.

FIGURE CAPTIONS

Fig. 1. (color online)

(a) Schematic of [001]$_C$-poled, "4R", domain-engineered PZN-4.5PT showing the <111>$_C$ polar vectors of its four constituent domain variants with respect to the poling direction; in the longitudinal mode, stress is applied along this axis. (b) After a stress-induced phase transition to an orthorhombic phase, four domain variants remain with polar vectors along <110>$_C$-type directions; the structure is no longer domain-engineered with respect to the poling direction. (c) [001]$_C$-poled PZN-4.5PT under transverse compressive stress. (d) In the resultant, stress-induced, orthorhombic phase, two domain variants remain; the structure is still domain engineered with respect to the poling direction.

Fig. 2. (color online)

Quasi-static, charge-stress loops for [001]$_C$-poled, "4R" domain-engineered PZN-4.5PT under (a) transverse and (b) longitudinal uniaxial, compressive stresses. In (b) the results of successive load/unload cycles are shown, each with increasing magnitude. The loops run anti-clockwise and clockwise in (a) and (b), respectively.

(over please)

Fig. 3. (color online)

Lattice strain, derived from X-ray diffraction measurements, at various static, uniaxial, compressive loads for [001]$_C$-poled, "4R" PZN-4.5PT. In (a) the stress is applied perpendicular to the poling direction (transverse mode) and the strain is measured along it; in (b) the stress is applied along the poling direction (longitudinal mode) and the strain is measured perpendicularly to it.

# REFERENCES

[1] S.-E. E. Park and W. Hackenberger, Current Opinion in Solid State and Materials Science **6,** 11 (2002).

[2] B. Noheda, Current Opinion in Solid State and Materials Science **6,** 27 (2002).

[3] S.-E. E. Park and T. R. Shrout, J. Appl. Phys. **82,** 1804 (1997).

[4] M. K. Durbin, E. W. Jacobs, and J. C. Hicks, Appl. Phys. Lett. **74,** 2848 (1999).

[5] B. Noheda, Z. Zhong, D. E. Cox, G. Shirane, S.-E. Park, and P. Rehrig, Phys. Rev. B **65,** 224101 (2002).

[6] H. Cao, J. Li, D. Viehland, and G. Xu, Phys. Rev. B **73,** 184110 (2006).

[7] C.-S. Tu, H. Schmidt, I.-C. Shih, and R. Chien, Phys. Rev. B **67,** 020102(R) (2003).

[8] M. Davis, D. Damjanovic, and N. Setter, J. Appl. Phys. **97,** 064101 (2005).

[9] M. Davis, D. Damjanovic, and N. Setter, Phys. Rev. B **73,** 014115 (2006).

[10] D. Viehland, J. Am. Ceram. Soc **89,** 775 (2006).

[11] M. Davis, D. Damjanovic, and N. Setter, J. Appl. Phys. **95,** 5679 (2004).

[12] Z. Feng, D. Lin, H. Luo, S. Li, and D. Fang, J. Appl. Phys. **97,** 024103 (2005).

[13] Q. Wan, C. Chen, and Y. P. Shen, J. Appl. Phys. **98,** 024103 (2005).

[14] C. S. Lynch, Acta Mater. **44,** 4137 (1996).

15  N. J. Ramer, S. P. Lewis, E. J. Mele, and A. M. Rappe, AIP Conference Proceedings **436,** 156 (1998).

16  G. Xu, H. Hiraka, G. Shirane, and K. Ohwada, Appl. Phys. Lett. **84,** 3975 (2004).

17  M. Davis, Thesis, Ecole Polytechnique Fédérale de Lausanne (EPFL), 2006.
13

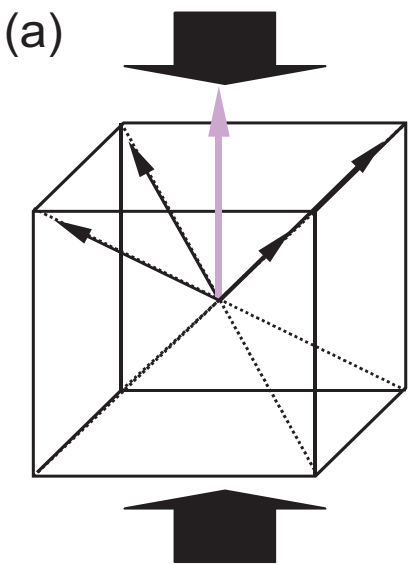 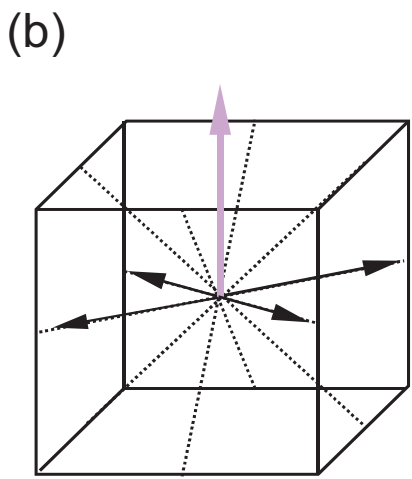
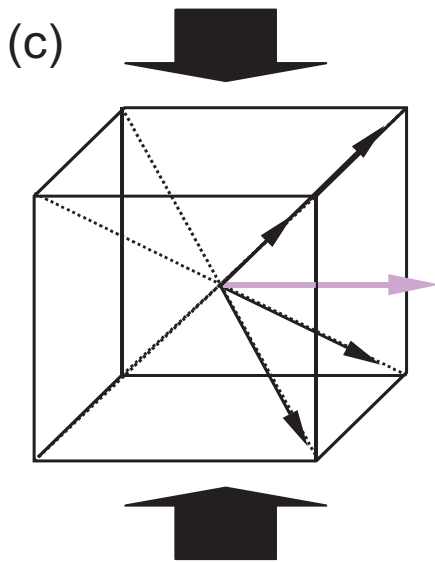 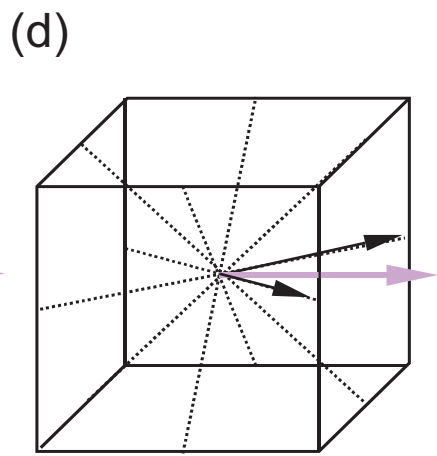

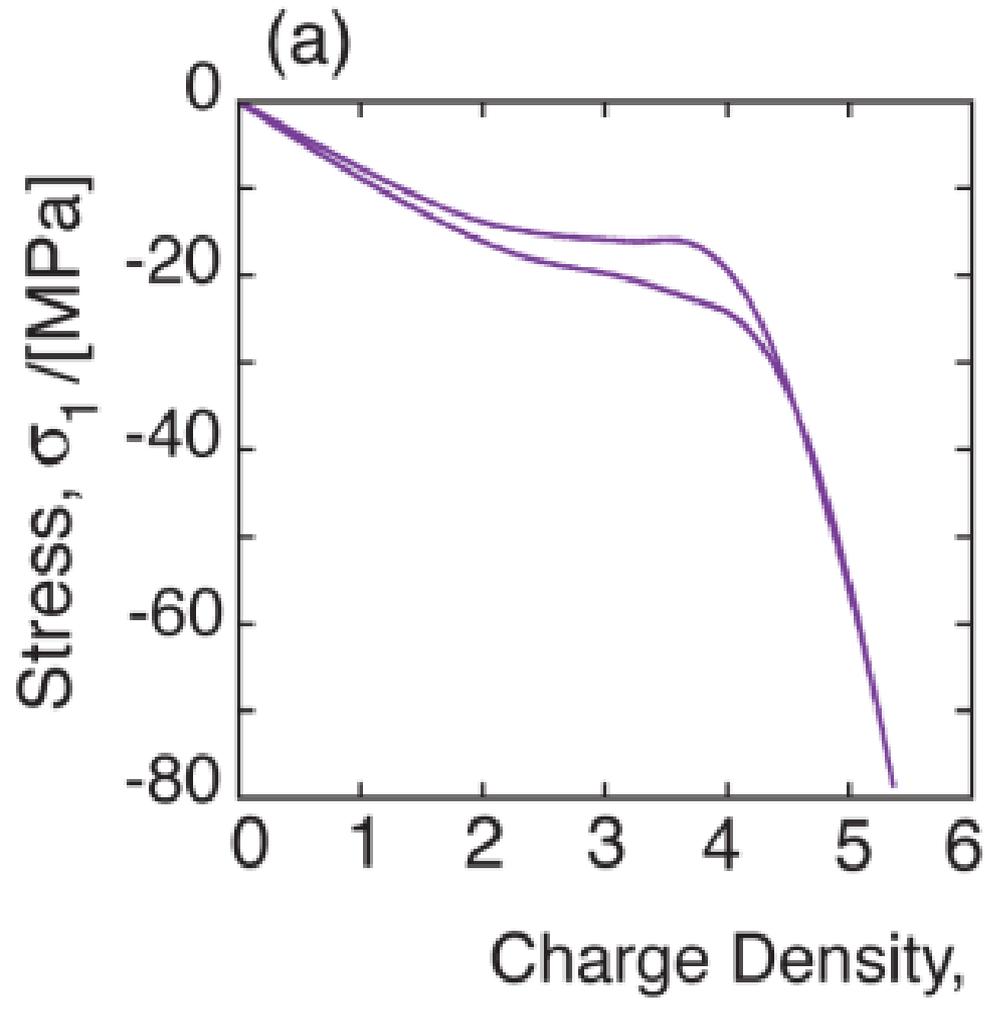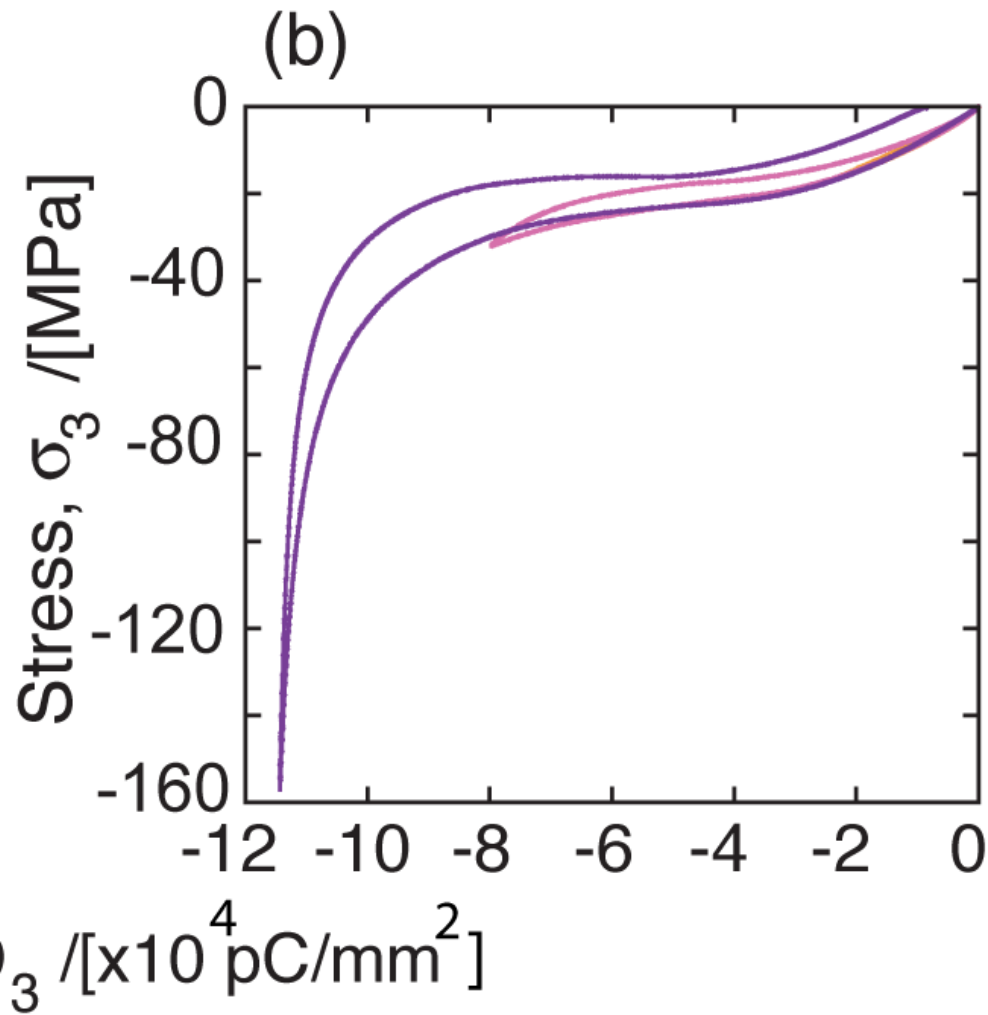

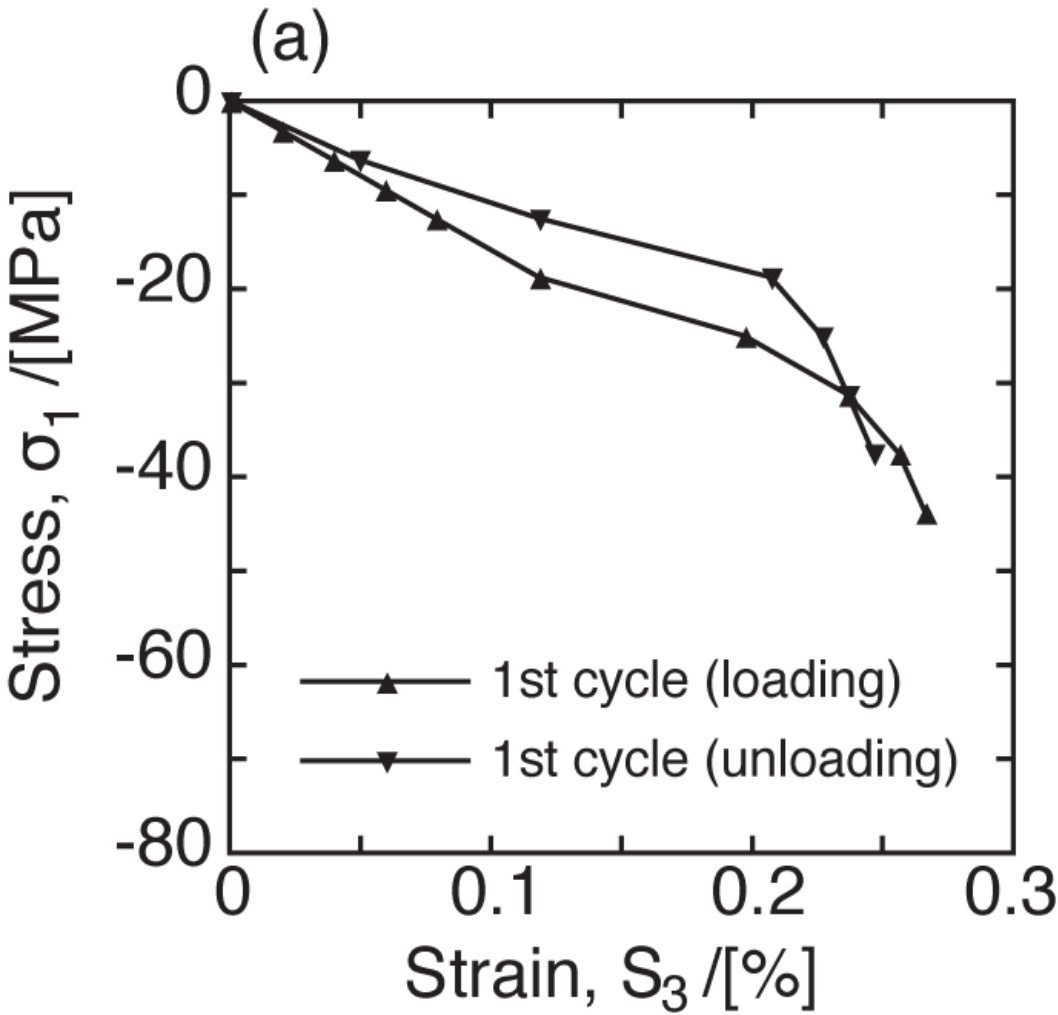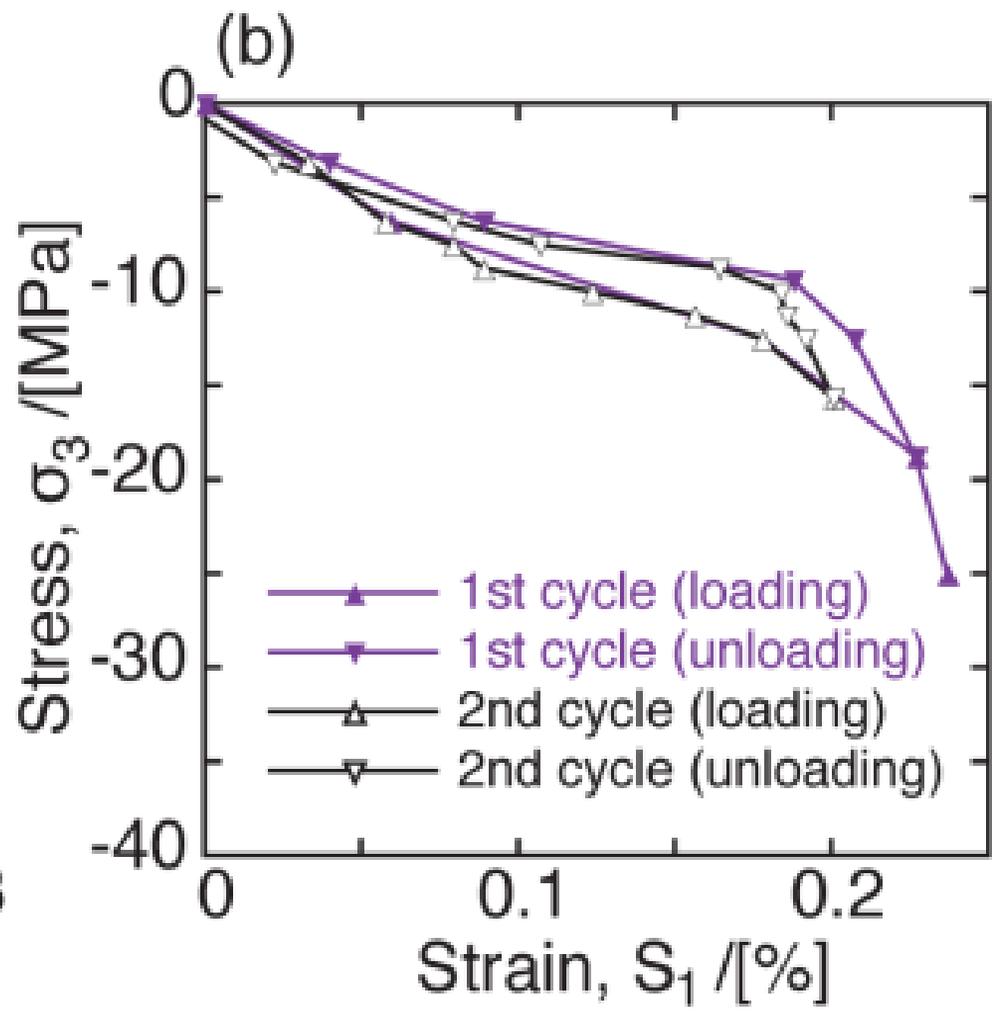